\documentclass{iaus}
\usepackage{graphicx}

\title[Magnetic fields in the descendants of massive OB stars] %% give here short title %%
{Searching for Magnetic Fields in the Descendants of Massive OB Stars}

\author[J.H. Grunhut et al.]   %% give here short author list %%
{J.H. Grunhut$^{1}$, G.A. Wade$^1$, D.A. Hanes$^1$, E. Alecian$^2$}

\affiliation{$^1$Kingston, Canada; $^2$Grenoble, France}

\pubyear{2010}
\volume{272}  %% insert here IAU Symposium No.
\pagerange{119--126}
\jname{Active OB stars}
\editors{A.C. Editor, B.D. Editor \& C.E. Editor, eds.}
\begin{document}

\maketitle

\begin{abstract}
We present the results of a recent survey of cool, late-type supergiants - the descendants of massive O- and B-type stars - that has systematically detected magnetic fields in these stars using spectropolarimetric observations obtained with ESPaDOnS at the Canada-France-Hawaii Telescope. Our observations reveal detectable, often complex, Stokes $V$ Zeeman signatures in Least-Squares Deconvolved mean line profiles in a significant fraction of the observed sample of $\sim$30 stars. 

\end{abstract}

\firstsection % if your document starts with a section,
\section{Introduction}
Supergiants are the descendants of massive O and B-type main sequence stars. Unlike their main sequence progenitors, cool supergiants are characterized by a helium-burning core and a deep convective envelope.

Due to their extended radii, low-atmospheric densities, slow rotation and long convective turnover times, supergiants provide an opportunity to study stellar magnetism at the extremes of parameter space.

In fact, observations of late-type supergiants show characteristics consistent with magnetic activity, such as luminous X-ray emission and flaring, and emission in chromospheric UV lines - phenomena suggesting the presence of dynamo-driven magnetic fields.

Motivated by the activity-related puzzles of late-type supergiants, the near complete lack of direct constraints on their magnetic fields, and recent success of measuring fields of red and yellow giants (e.g. Auri\`{e}re et al. 2008), we have initiated a program to search for direct evidence of magnetic fields in these massive, evolved stars. Here we summarize the recent results of Grunhut et al. (2010).

\section{Observations}
Circular polarization (Stokes $V$) spectra were obtained with the high-resolution (R$\sim$68000) ESPaDOnS and NARVAL spectropolarimeters at the Canada-France-Hawaii Telescope and Bernard Lyot Telescope, as part of a large survey investigating the magnetic properties of late-type supergiants.

To date, we have observed more than 30 stars: 4 A-type stars, 8 F-type stars, 11 G-type stars, 7 K-type stars, and 3 M-type stars.

\section{Magnetic Field Diagnosis \& Results}
We applied the Least-Squares Deconvolution (LSD; Donati et al. 1997) technique to all our data in order to increase the S/N and detect weak Zeeman signatures.

\begin{figure} 
\centering
\includegraphics[width=3.6in]{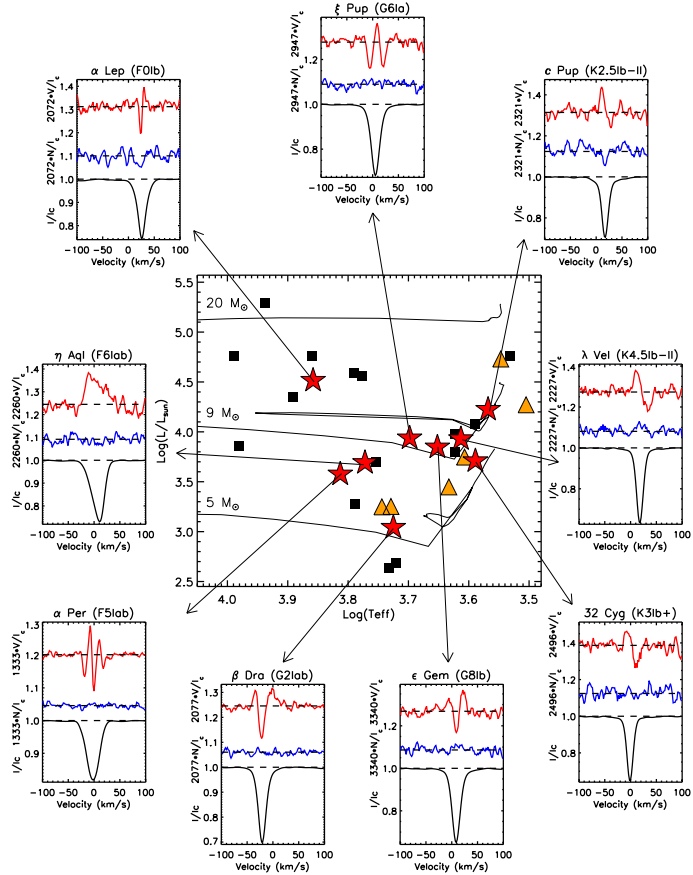}
%~~~\includegraphics[width=1.45in]{lsd_grid1.eps}
\caption{HR diagram showing all observed supergiants. Black squares indicate stars for which no Zeeman signatures were detected, orange triangles indicate stars with suggestive Zeeman signatures, while red stars represent the supergiants with clear Zeeman signatures. Surrounding the HR diagram are illustrative mean Stokes $V$ (top), diagnostic null (middle), and unpolarized Stokes $I$ (bottom) LSD profiles of the 9 stars with clear detections.}
\label{lsd}
\end{figure}

In Fig.~\ref{lsd} we present all stars with clear Zeeman signatures detected in Stokes $V$. Also shown in Fig.~\ref{lsd} is the placement of all observed stars on an HR diagram.

Our investigation shows that many late-type supergiants host detectable Stokes $V$ Zeeman signatures, which are frequently complex. Overall, we find that approximately 1/3 of our sample reveal detectable Zeeman signatures in Stokes $V$. However, we find no clear differences between classical activity indicators (such as Ca\,{\sc II} H\&K or H$\alpha$ emission) of those stars with or without detections. However, we do find a weak correlation between the Ca\,{\sc II} core equivalent width and the magnetic field strength for those stars with multiple observations. In addition, we also see clear temporal variability of the Stokes $V$ profiles for those targets with multiple observations.

\end{document}